# Self-blocking of interstitial clusters near metallic grain boundaries

Xiangyan Li[1], Wei Liu[1], Yichun Xu[1], C.S. Liu[1*], B.C. Pan[2*], Yunfeng Liang[1,3], Q.F. Fang[1], Jun-Ling Chen[4], G.-N. Luo[4], Zhiguang Wang[5], Y. Dai[6]

[1]Key Laboratory of Materials Physics, Institute of Solid State Physics, Chinese Academy of Sciences, P.O. Box 1129, Hefei 230031, PR China,

[2]Hefei National Laboratory for Physical Sciences at Microscale and Department of Physics, University of Science and Technology of China, Hefei 230026, PR China,

[3]EnvFement and Resource System Engineering, Kyoto University, Kyoto 615-8540, Japan,

[4]Institute of Plasma Physics, Chinese Academy of Sciences, Hefei 230031, PR China,

[5]Institute of Modern Physics, Chinese Academy of Sciences, Lanzhou 730000, PR China,

[6]Spallation Neutron Source Division, Paul Scherrer Institut, 5252 Villigen PSI, Switzerland.

[*]Corresponding author. Tel.: +86 0551 65591062. E-mail: csliu@issp.ac.cn

Nano-crystallize materials have been known for decades to potentially owe the novel self-healing ability for radiation damage, which has been demonstrated to be especially linked to preferential occupation of interstitials at grain boundary (GB) and promoted vacancy-interstitial annihilation. A major obstacle to better understanding the healing property is the lack of an atomistic picture of the interstitial states near GBs, due to severely separation of the timescale of interstitial segregation from other events and abundance of interstitials at the GB. Here, we report a generic "self-blocking" effect of the interstitial cluster ($SIA_n$) near the metallic GB in W, Mo and Fe. Upon creating a $SIA_n$ near the GB, it is immediately trapped by the GB during the GB structural relaxation and blocks there, impeding GB's further spontaneous trapping of the $SIA_n$ in the vicinity and making these $SIA_n$s stuck nearby the GB. The $SIA_n$ in the stuck state surprisingly owes an exceptionally larger number of annihilation sites with vacancies near the GB than the $SIA_n$ trapped at the GB due to maintaining its bulk configuration basically. Besides, it also has an unexpectedly long-ranged repelling interaction with the $SIA$ in the bulk region, which may further affect the GB's trap of the $SIA_n$. The self-blocking effect might shed light on more critical and extended role of the GB in healing radiation-damage in NCs than previously recognized the GB's limited role based on GB's trap for the $SIA$ and resulted vacancy-$SIA$ recombination.

Our familiar metals and alloys in daily lives are all poly-crystalline materials, where grain boundaries (GBs) generally play a role in mechanical properties like the well-known Hall-Petch-type behavior of the apparent strength and grain size and mass transport like reduced electrical conductivity due to the blocking effect of GBs []. In the past two decades, emerging experimental evidence suggests that GBs also have an important role in healing radiation-damage of nano-crystallize materials (NCs), which has drawn considerable attention in nuclear materials fields in recent years [9,10]. The micro-mechanism of the self-healing property is unfortunately still poorly understood.

Under irradiations of neutrons, protons or heavy ions, common large-grained materials are damaged through the creation of point defects i.e. self-interstitial atoms (*SIA*s) and vacancies, and clusters including bubbles, voids and dislocation loops. Yet, it is conventionally thought as an unsurprising fact for NCs to exhibit self-healing ability, since GBs prevalent in NCs are good sinks for all kinds of defects and naturally trap radiation-created defects, consequently healing radiation-damage in NCs. This explanation requires activation of defects motion in the grain, particularly vacancies that have low mobility, which nevertheless conflicts some experimental results: when the temperature is too low to stimulate single vacancies diffusion in the grain, the self-healing property of NCs, however, still remains. It seems that NCs may intrinsically possess the radiation-damage healing ability, which has been demonstrated to be relevant to the GB-enhanced vacancy diffusion in recent 5 years. Meanwhile, another role of the GB as the catalyzer for vacancy-*SIA* annihilation is revealed: the *SIA* preferentially segregating into the GB is known to recombine the vacancy nearby the GB at a low energy barrier, truly eliminating immobile vacancies []. Quite recently, it is proposed that the *SIA* within the GB also causes GB's motion or phase transition, contributing to defects elimination [].

Among all the events involved in the GB-enhanced vacancy diffusion and promoted vacancy-*SIA* recombination, the *SIA* segregation may play a outstanding role due to its severely separation of the timescale from other events. The *SIA* has exceptionally high mobility (at room temperature, the vacancy with a barrier of 1.8 eV in W bulk takes about $10^{18}$ second to jump one step, while the *SIA* with a barrier of 0.02 eV only takes about $10^{-12}$ second to migrate one step). Radiation-created *SIA*s will segregate into the GB far earlier than vacancies, which in turn modify local GB structure and change vacancy behaviors of diffusion and annihilation. In addition, *SIA*s will reside at the GB in significant concentrations. As revealed by molecular statics (MS) calculations, only several atomic layers near the GB serve as storage sites for *SIA*s. These layers take a small volume fraction of only about 3% even for the grain size of 100 nm. Assuming $C_{bulk}$ to be concentration of *SIA*s in the bulk, as all *SIA*s locate at the GB, the *SIA* concentration at the GB will be 97%/3%*$C_{bulk}$, which is 32.5 times $C_{bulk}$. This indicates that *SIA*s will be rather rich near the GB. Therefore, a major obstacle to better understanding the healing property is the lack of an atomistic picture of the *SIA*s states near GBs.

Following on from our recent investigations of fundamental interactions of the single vacancy and *SIA* with the GB in W, Mo and Fe, which have revealed that the GB has a positive but limited role in removing vacancies nearby, we report a molecular statics (MS) study of the important *SIA* cluster ($SIA_n$) behaviors near the GB. Here, we show a generic "self-blocking" effect for the $SIA_n$ near the GB in the investigated W, Mo and Fe systems. A $SIA_n$ that is initially created near the GB is instantly trapped by the GB during the GB structural relaxation and blocks there, which then hinders GB's further spontaneous trapping of the $SIA_n$ in the vicinity and makes these $SIA_n$s stuck nearby the GB. The $SIA_n$ in the stuck state surprisingly maintains its bulk configuration basically and consequently owes an exceptionally larger number of annihilation sites with vacancies near the GB than the $SIA_n$ trapped at the GB. Meanwhile, it also has an unexpectedly long-ranged repelling interaction with the *SIA* in the bulk region, which may further affect the GB's trap of the $SIA_n$. This self-blocking of the $SIA_n$ near the GB provides a unique spatially-efficient route to eliminating bulk vacancies and give new atomic insights into the self-healing ability of nano-crystallize materials.

# Results

**Trapping of interstitials clusters by the GB.** As this work focus on $SIA_n$ behaviors near the GB, We first determine ground states for the $SIA_n$ in the bulk region far away from the GB by adopting the "constraint-search" method (see Methods section for "Search of stable configuration of bulk interstitial clusters"). Interatomic interactions are described by empirical embedded-atom method potentials []. Note that such configuration exploring is obvious beyond the ability of first-principles calculations due to requirement of a large model size and a large configurational space (particularly for the $SIA_n$ in W and Mo). For example, in two body-centered cubic lattice cells, the cluster $SIA_5$ has about $C_{14}^5 \times 2^5 = 64064$ combinations (each $SIA$ in $SIA_5$ is assigned to two adjusting positions that are $\pm 1/4<111>$, leading to $2^5$ combinations).

Figure 1 presents the obtained configurations for the interstitial clusters $SIA_1$–$SIA_{10}$ in W bulk, where one <111> crowdion in the $SIA_n$ is indicated by three green spheres lined together (in reality, one $SIA$ is composed of about 7 displaced atoms that are chained along <111> direction within a range of 6 nearest distances). It can be seen the stable configurations of small interstitial clusters are composed of compact <111> crowdions parallel to each other (see Supplementary Fig. **S1** for $SIA_n$ configurations characterized by potential energy contour for interstitial clusters $SIA_1$–$SIA_{10}$ in W bulk). Besides, the crowdions in the $SIA_n$ are tightly bound to the $SIA_n$ due to large binding energies of 3–6 eV (see Supplementary Fig. **S2** for the $SIA_n$ formation energies and binding energies of one $SIA$ to the $SIA_n$ in W bulk). Thus, the $SIA$ in the bulk indeed has a strong tendency to be clustered, which rationales our investigation of interstitial clusters behaviors near the GB rather than that of the single interstitial as considering the radiation-damage healing issue. In addition, large amounts of meta-stable configurations exist, indicating complexity of the configurational space for the $SIA_n$. For example, the energy for the meta-stable configuration "O-;I+" for $SIA_2$ is just 0.3 eV higher than the stable one "O-;I-". $SIA_3$ has 9 meta-sable configurations with energy deviation from the stable one less than 1 eV.

With above bulk configurations of the $SIA_n$ in hand, behaviors of the $SIA_n$ near the symmetric tilt GB $\Sigma 5(3\ 1\ 0)/[0\ 0\ 1]$ in the BCC metal W are investigated used molecular statics (MS). Through putting one $SIA_n$ ($n$ ranging from 1 to 10) near the GB (see Methods section for "Put of bulk interstitial clusters near the GB") and relaxing the GB system, we calculate the $SIA_n$ formation energy, the energy indicating its thermal stability, within the range of about 15 Å from the GB. Consequently, the energy landscape of the $SIA_n$ near the GB is obtained. For the case of the single interstitial, the landscape is given in Fig. 2a as an example, which is characterized by contour of the $SIA$ formation energy near the GB in Supplementary Fig. **S3**. It can be seen, the landscape exhibits several features. First, we note that the $SIA$ formation energy is greatly reduced as one $SIA$ approaches to the GB core from the bulk region (*Bulk region* in Fig. 2a), passing through a transition region (*Transition region* in Fig. 2a) where the $SIA$ formation energy shows little deviation from the bulk value. As the $SIA$ locates at the center of *C1E1*, its energy is as low as 2.32 eV and the reduction is as large as about 7.16 eV relative to the bulk value of 9.48 eV. Here, the GB sink strength for the single $SIA$ calculated by an EAM potential [] is consistent with our previous results of about 7.5 eV, which is calculated through a BOP potential []. The reduction, defined as the binding energy of the $SIA_n$ to the GB, is even larger for the $SIA_2$–$SIA_{10}$, which basically increases linearly with the number of interstitials $n$ in the cluster as shown in Fig. 2b. The exceptionally large binding energy can be ascribed to considerably lower $SIA_n$ formation energy at the GB than that in the bulk (see (Fig. 2b) and Supplementary Fig. S2). These energetic calculations not only indicate a strong thermodynamic driving force for the $SIA_n$ to move from the bulk to the GB, but also suggest an exceptionally strong binding of the $SIA_n$ to the GB, which clearly defines the GB's trapping for the $SIA_n$ (*Trapping region* in Fig. 2a). By monitoring and visualizing the GB structural relaxation process as calculating $SIA_n$ formation energies, we find that, the $SIA_n$ that is initially created at the site of 5–6 atomic layers near the GB core with very low formation energies will spontaneously flows into the GB during the relaxation (see Supplementary Figs. S4 and S5 for the trapping process of the *SIA* and $SIA_3$). Therefore, the GB's trapping for the $SIA_1$–$SIA_{10}$ is a spontaneous process, although these defects may have different dynamic behaviors in the bulk that lie outside the scope of this paper. Furthermore, we find that only 2–3 layers serve as storage sites for the $SIA_n$ by examining the exact location of the $SIA_n$ at the GB core after it is trapped there.

Figure 2c vividly presents this GB's role as a trap for the $SIA_n$ where $SIA_n$ formation energies near the GB are normalized by dividing respective bulk values. The curves of the $SIA_n$ formation energies with distance to the GB look like a "squared potential well" with a depth about 1-0.4=0.6. This value is consistent with our previous calculations of the single *SIA* near several GBs in W, Mo and Fe [], indicating a common feature in the GB's trapping for all the $SIA_n$, although they have different formation energies at the GB prior to normalization. Meanwhile, the scattered feature is observed in the $SIA_n$ formation energies at the GB core, which agrees well with the anisotropy in energy landscape. As shown in Fig. 2a, when an *SIA* moves from *G1* to *D2* though the path of *G1-F1-E1-D2*, the energies on this path are 9.40, 2.28, and 2.70 eV, respectively. While it diffuses from another point *G2* that has the same distance normal to the GB as *G1*, the energies on the path of *G2-E1-C1* are 2.32 and 2.32 eV, correspondingly. The anisotropy is also clearly presented in Supplementary Fig. **S3**. The anisotropy in the energy landscape implies that, the exact location of the *SIA* after it is trapped at the GB core depends on its initial position and orientation near the GB.

To better understand the extremely tight trapping of the $SIA_n$ by the GB, we further give the trapped states of the $SIA_1$–$SIA_{10}$ at the GB by visualizing the GB structure that captures a $SIA_1$–$SIA_{10}$. In Fig. 3, such visualization is implemented both in real space and potential energy space. In real space (Fig. 3a), it can be seen the $SIA_1$–$SIA_5$ reside at the GB severely locally, behaves like points with the single $SIA$ exactly like a point and $SIA_2$–$SIA_5$ consisting of several points. The $SIA_6$–$SIA_{10}$, however, cannot be clearly presented in real space, although the rough and rugged stress fields around them are also observed similar to limited stress fields surrounding the $SIA_1$–$SIA_5$. The single $SIA$ and clusters basically lie within the limited GB stress field with semi-width of about 5 Å. In potential energy space (Fig. 3b), the local point-like states for the $SIA_1$–$SIA_{10}$ are distinctly presented. This "point-like" state for $SIA_1$–$SIA_{10}$ at the GB is severely different to that in the bulk which composes of bunches of parallel <111> crowdions (see Fig. 1 and Supplementary Fig. **S1**). Besides for difference in morphologies, the stress fields around interstitials at the GB are also evidently different to that in the bulk. The elongated shape of the single $SIA$ in W has an ellipsoid stress field around the center of the SIA with the semi-major axis is 18 Å in <111> direction and semi-minor axis 5.4 Å) []. This unique feature of interstitials at the GB in both morphologies and ranges of stress fields may result in an exceptionally large release in volume distortion energy from bulk bunches of <111> crowdion.

About the trapping of the $SIA_n$ by the GB, we conclude that, the GB spontaneously traps the $SIA_n$ lying within the range of about 10 Å of the GB (Fig. 2), but stores the $SIA_n$ at the sites of about 5 Å from the GB (Fig. 3). The $SIA_n$ is tightly bound to the GB. These conclusions do not violate our intuition and expectation of the GB's role as defects sink. How do these interstitials trapped at the GB modify behaviors of defects near the GB? At least the vacancy occupation and interstitial morphology will be modified. (i) Previously, Bai et al has shown that the single $SIA$ at the GB in Cu significantly influences the diffusion and annihilation of the vacancy near the GB []. So is true in W, Mo and Fe systems [], although the influence is local in all these systems. (ii) As shown in Fig. 3b, it seems the GB does not completely absorb $SIA_7$–$SIA_{10}$, with parts of $SIA$s in these clusters being stuck near the GB. This hints that the $SIA_n$ may maintain its bulk morphology near the GB. The possibility is seen for the bulk SIA, which still possesses a perfect <111> crowdion after its trapping by one $SIA_n$ in the bulk with the binding energy at the level of 3–6 eV (see Supplementary Figs. **S1** and **S2**). Here, our attention is particularly paid to the $SIA_n$ morphology near the GB because of its possible relation to the GB healing ability for radiation damage as described and discussed in the following paragraphs.

**Self-blocking of interstitial clusters near the GB.** To investigate the effect of the $SIA_n$ trapped at the GB on newly-coming $SIA_n$s from the bulk (the $SIA_n$ at different sites from the GB will arrive at the GB at time $t=L^2/6D$ where $L$ is the average diffusion length and $D$ is diffusion coefficient), near the GB in Fig. 3a, we add more $SIA_n$s (see Methods section for "Put of bulk interstitial clusters near the GB"). The sites are chosen at the border of the potential well (as marked in Fig. 2). After fully relaxing the GB system, we examine the $SIA_n$ states. To our big surprise, the newly added $SIA_n$s no longer spontaneously flow into the GB, but reside near the GB. Figure 4 presents such state. We can see, the newly put $SIA_n$s basically maintain their shape of <111> crowdion with one end of their configurations bound to the GB. Meanwhile, the intrinsic large range of stress fields around the $SIA_n$ naturally remains, which can extend as far as 20 Å into the bulk region and is four times half-width of the GB that is pure or with $SIA_n$s. This difference in morphologies is also embodied in the formation energies of the $SIA_n$. For the $SIA_n$ trapped at the GB and in the bulk, they have two rather different levels of formation energies that show linear increase with the number of $SIA$ in the cluster $n$ (Fig. 5a). For the $SIA_n$ near the GB in Fig. 4, their formation energies, however, fluctuate within these two levels (Fig. 5a), indicating that these $SIA_n$s are actually stuck in the vicinity of the GB.

Particularly, the $SIA_n$ that well maintains its bulk configuration has energy comparable to that of the bulk one. For example, the $SIA_1$–$SIA_{10}$ stuck near the GB/in the bulk have separate energies of 8.84/9.48, 16.41/16.63, 21.21/22.60, 26.00/27.87, 31.63/33.28, 38.50/38.15, 42.88/41.69, 48.84/46.40 and 57.54/53.97 eV. We also note that, the energies of some $SIA_n$s are slightly higher as they are stuck near the GB than in the bulk like the $SIA_6$–$SIA_{10}$ (see Supplementary Figs. S6 and S7 for the energetic feature and electron charge distribution of the $SIA_n$ stuck near the GB together with that in the bulk and trapped at the GB).

Due to blocking of the $SIA_n$ at the GB and occupying vacant space there, the GB's trapping for the $SIA_n$ near itself is thus hindered; we refer to this effect as the self-blocking effect of interstitial clusters near the GB. In the self-blocking effect, $SIA_n$s that are cruise in the bulk are pinned to the neighboring hood of the GB, however, still maintaining their bulk configurations due to blocking. To understand how this effect affects the vacancy annihilation, we calculate vacancy-$SIA_n$ annihilation volume to measure the annihilation probability. We first calculate the vacancy formation energy of the sites around the $SIA_n$ (see Methods section for "Calculation of defects energetic properties"). The volume is defined as the number of sites with negative formation energies; as we create one vacancy at this type of sites, it is spontaneously recombined. More information is referred to Supplementary Fig. **S8** where $SIA_n$ configurations are characterized by vacancy formation energy around the $SIA_n$ in W bulk, trapped at the GB, and stuck near the GB. Figure 5b presents the obtained annihilation volume for the $SIA_n$ stuck near the GB and results for the $SIA_1$–$SIA_{10}$ in the bulk and trapped at the GB are also given for comparison. We find that, a $SIA_n$ stuck near the GB has the largest annihilation volume which is far larger than that for the $SIA_n$ trapped at the GB. For example, the volume for the single $SIA$ is as particularly large as 60, which is, however, only 3 for the trapped one at the GB. Although the interstitials in the bulk have the annihilation volume comparable to that of blocking ones, in large-grained materials, they often easily flow to surfaces [], having no role in removing vacancies. These calculations indicate that interstitials states strongly influence their probabilities for annihilating vacancies. The $SIA_n$ stuck near the GB owes strong abilities for removing vacancies nearby.

We note that, as the $SIA_n$s are stuck near the GB with one end of their configurations bound to the GB, they have a cross section fronting the bulk region. The properties of this cross section are investigated by calculating binding energies of one single $SIA$ to $SIA_1$–$SIA_{10}$ in <111> direction. As shown in Fig. 5c and 5d, we surprisingly find that the binding energies are negative, indicating repelling interaction of the $SIA_n$ with the $SIA$ in <111> direction which is basically enhanced with increasing the number of interstitials in the cluster (Fig. 5d). Besides, the interaction range is rather large, which extends to more than 20 Å.

Figure 6 sums up and illustrates fundamental pictures about interstitials states near the GB in W derived from our above energetic calculations of the $SIA_n$. The GB acts as a squared potential well for the $SIA_n$, spontaneously trapping $SIA_n$ near itself. In the bulk, the $SIA_n$ has large stress fields surrounding bunches of <111> crowdions, whereas behaving like points as trapped at the GB. These interstitials, blocking at the GB, hinder the GB's further trapping of the $SIA_n$ and make newly arriving $SIA_n$ to be stuck near the GB. The $SIA_n$ in different states has respective "energy levels". For the stuck one, its energy $E_2$ is less than bulk value $E_1$, but higher than the energy of the trapped one at the GB $E_3$. The interstitial with energy higher than $E_1$ has a thermodynamic trend to diffuse away from the GB. The $SIA_n$ stuck near the GB not only has large annihilation volume for bulk vacancies, but also owes one repelling cross section in the direction along <111>.

## Discussion

To demonstrate the generality of this "self-blocking" effect as observed near a W GB, we investigate $SIA_n$ behaviors near the symmetric tilt GB $\Sigma 5(3\ 1\ 0)/[0\ 0\ 1]$ in Mo that have similar defects properties to W (see Supplementary Figs. **S9**–**S12**). The conclusion is drawn similar to W. For example, stable $SIA_1$–$SIA_{10}$ in Mo bulk are found to be composed of bunches of <111> crowdions. The $SIA_n$s are trapped by the GB as we put them near the GB. The trapped $SIA_1$–$SIA_6$ owe local point-like configurations with exceptionally low energies at the GB compared with bulk values (1.73/7.37, 2.97/12.82, 4.60/17.68, 8.19/21.73, 6.28/25.77, and 9.05/29.60 eV for $SIA_1$–$SIA_6$ respectively). The $SIA_n$ stuck near the GB has larger annihilation volume surrounding itself than that in the bulk and trapped at the GB.

One more BCC system, Fe that with unique stable configuration of <110> dumbbell for its single interstitial in the bulk, is investigated (see Supplementary Figs. **S13**–**S16**). Similar phenomenon is also observed to W and Mo. But, big difference exists. Calculations show that, the most favorable configurations for $SIA_1$–$SIA_3$ are made of perfect parallel <110> single dumbbells, consistent with first-principle results []. The configurations for $SIA_4$–$SIA_6$, however, are found to be complex, looking like of compact deformed <110> dumbbells. The $SIA_7$–$SIA_{10}$ has similar configurations of parallel <111> crowdions to that of W and Mo. As we put one *SIA* cluster nearby the GB, it is instantly trapped by the GB. During the following adding of interstitials, self-blocking effect can only be observed as the adding lasts 4–8 times for $SIA_1$–$SIA_7$ and 2–3 times for $SIA_8$–$SIA_{10}$, which is obviously dissimilar to that in W and Mo where the blocking is observed just after one more interstitial or cluster is added to the site nearby the trapped $SIA_n$. Some deep and rooted reasons may be related to interstitials configurations and their diffusion mechanism. In W and Mo, interstitials diffuse along <111> direction one-dimensionally, while in Fe interstitial migrates via well-known translation-rotation Johnson's mechanism []. By examining the absorption process, we show that, as the <111>-type $SIA_n$ is absorbing, displaced atoms along <111> that compose the $SIA_n$ gradually recover to the normal lattice sites, while the front atom of the $SIA_n$ gets trapped at the GB. This atom can block at the GB and impede further the GB's further trapping of <111> crowdions near the GB. For <110> dumbbell, we find, it can be trapped by the GB through several different curved chains of atoms, rather than the sole <111> linear chain of atoms for <111> crowdion. In potential energy space (Supplementary Figs. 15 and 16), the displaced atoms near the GB gradually get isolated and undergo "bond-breaking". In the end, the displaced atoms recover to their lattice sites. The dumbbell disappears and the inserted atom gets trapped at the GB like a point. To our knowledge, this self-blocking effect for interstitial clusters is not directly observed experimentally. Interestingly, an analogy phenomenon exists where the occupation of one type of impurities near the dislocation is influenced by another one that preferentially locates at the dislocation []. Experiments show that the segregation of hydrogen to dislocations at 323 K decreases markedly in palladium doped with interstitial carbon or boron after a relatively low temperature annealing, 423 K. The blocking of heavy interstitial atoms, carbon or boron at dislocations is attributed to reduced hydrogen occupation at the GB [].

In conclusion, using atomistic simulations, we demonstrate that radiation-created interstitial clusters can be stuck near the GB due to blocking of the trapped $SIA_n$ at the GB. This self-blocking effect is generically observed near a symmetric tilt GB in W, Mo and Fe. Previously [], the GB's role in healing radiation damage is mainly attributed to its preferential absorption of interstitials and low-barrier annihilation of the vacancy around the interstitial. However, based on this picture, the role is local although working with kinetic efficiency. In the self-blocking effect, the $SIA_n$ stuck near the GB, maintaining its bulk configuration basically, provide an exceptionally large number of annihilation sites for bulk vacancies than trapped ones at the GB that behave like a point. The self-blocking of interstitials near the GB may not only give new atomic insights into self-healing in nano-crystallize materials, but also provide a novel and efficient approach to eliminating bulk vacancies: the defects that pin radiation-created $SIA_n$ and do not modify $SIA_n$ configurations will be decorated as good agents for removing vacancies.

Alloying elements and impurities may serve as this type of defects together with our investigated GBs.

## Methods

**Interatomic potentials and calculation models in MS calculations.** Molecular statics (MS) simulations were performed to calculate interstitials formation energy and binding energies in the bulk and near the GB. The embedded-atom-method (EAM) inter-atomic potentials were used to describe the interatomic interactions in W and Mo [29], and Fe [28]. Our test showed that these potentials accurately reproduces experimental values of the equilibrium lattice constant, cohesive energy, elastic constants, vacancy (interstitial) formation energy and migration energy in the lattice, and other properties of these investigated systems. The GB model is an symmetric tilt $\Sigma 5$ (3 1 0)/[0 0 1]. Its construction and relaxation can be found elsewhere []. Here the system to be studied is based on its engineering importance: bcc Fe is the matrix structure of ferritic/martensite steels which are considered to be the most promising structural materials in nuclear reactors. Mo and W are promising for usage as plasma facing components in the International Thermonuclear Experimental Reactor (ITER). These materials are exposed to severe irradiation environments.

**Search of stable configuration of bulk interstitial clusters.** Two constraints were imposed on exploring the configuration space of interstitial-type defects: (1) the single SIA composing the cluster is <111> crowdion in W and Mo, which is <110> dumbbell in Fe; (2) for $SIA_2$-$SIA_5$, search is performed only within one lattice cell, while search is conducted within two lattice cells for the cluster $SIA_6$-$SIA_{10}$. The initial configuration of one SIA in W and Mo is created through inserting one atom at the lattice $\pm 1/4$<1 1 1>, and at the lattice $\pm 1/4$<1 1 0>. If the lattice is O point, and the SIA has its stable configuration at O-$1/4$<1 1 1>, then the configuration is recorded as O-. This is important when reproducing such stable configurations like near the GB.

**Put of bulk interstitial clusters near the GB.** A reference lattice point is selected near the GB as the one in a lattice cell like point O in Fig. 1. Then, the two lattice cells in Fig. 1 are transformed from the [100], [010] and [001] coordinate system to the one in the GB coordinate system [001], [1$\bar{3}$0], and [310]. The initial configuration for the $SIA_n$ near the GB is thus created using above notations for bulk $SIA_n$.

**Calculation of defects energetic properties.** Formation energy for the interstitial cluster $SIA_n$ in the bulk is defined as $Ef_{SIAn}^{bulk} = E_{SIAn}^{bulk} - E_{bulk} - nE_{coh}$, where $E_{SIAn}^{bulk}$ and $E_{bulk}$ are the total energies of the bulk simulation cell with and without the $SIA_n$, respectively. $E_{coh}$ is the cohesive energy per atom of a perfect bcc lattice, which is -8.90, -6.82, and −4.12 eV for W, Mo, and Fe, correspondingly. The binding energy of one SIA to the $SIA_n$ is calculated as $Eb_{SIA-SIAn}^{bulk} = Ef_{SIAn+1}^{bulk} - Ef_{SIAn}^{bulk} - Ef_{SIA}^{bulk}$.

For the SIAn at the site $\alpha$ near the GB, its formation energy is defined as $Ef_{SIAn}^{GB,\alpha} = E_{SIAn}^{GB,\alpha} - E_{GB} - nE_{coh}$, where $E_{SIAn}^{GB,\alpha}$ and $E_{GB}$ are the total energies of the GB system with and without the $SIA_n$ at the site $\alpha$, respectively. The binding energy of the $SIA_n$ to the site $\alpha$ at the GB is defined as $Eb_{SIAn-GB}^{GB} = Ef_{SIAn}^{bulk} - Ef_{SIAn}^{GB,\alpha}$.

When calculating vacancy-$SIA_n$ annihilation volume, we need to know vacancy formation energy at the site $\alpha$ around the $SIA_n$ that is in the bulk, trapped at the GB, or stuck near the GB. Vacancy formation energy is defined as $Ef_V^{\alpha} = E_V^{\alpha} - E + E_{coh}$, where $E_V^{\alpha}$ and $E$ are the total energies of the system with and without one vacancy at the site $\alpha$, respectively.

References

1. Rose, M., Balogh, A.G. & Hahn, H. Instability of irradiated induced defects in nanostructured materials. *Nucl. Instr. and Meth. B* **127**, 119-122 (1997).

2. Chimi, Y. *et al.* Accumulation and recovery of defects in ion-irradiated nanocrystalline gold. *J. Nucl. Mater.* **297**, 355-357 (2001).

3. Kurishita, H. *et al.* Development of ultra-fine grained W-TiC and their mechanical properties for fusion applications. *J. Nucl. Mater.* **367-370**, 1453-1457 (2007).

4. Kurishita, H. *et al.* Development of ultra-fine grained W-(0.25-0.8)wt%TiC and its superior resistance to neutron and 3 MeV He-ion irradiations. *J. Nucl. Mater.* **377**, 34-40 (2008).

5. Kilmametov, A.R., Gunderrov, D.V., Valiev, R.Z., Balogh, A.G. & Hahn, H. Enhanced ion irradiation resistance of bulk nanocrystalline TiNi alloy. *Script. Mater.* **59**, 1027-1030 (2008).

6. Kilmametov, A. *et al.* Radiaiton effects in bulk nanocrystalline FeAl alloy. *Radiat Eff. Defects Solids.* **167-8**, 631-639 (2012).

7. Yu, K.Y. *et al.* Radiation damage in helium ion irradiated nanocrystalline Fe. *J. Nucl. Mater.* **425**, 140-146 (2012).

8. Sickafus, K.E. *et al.* Radiation tolerance of complex oxides. *Science* **289**, 748-751 (2000).

9. Grimes, R.W., Konings, R. J.M. & Edwards, L. Greater tolerance for nuclear materials. *Nat. Mater.* **7**, 683-685 (2008).

10. Ackland, G. Controlling radiation damage. *Science* **327**, 1587-1588 (2010).

11. Odette, G.R., Alinger, M.J. & Wirth, B.D. Recent developments in irradiation-resistant steels. *Annu. Rev. Mater. Res.* **38**, 471-503 (2008).

12. Yvon, P. & Carre, F. Structural materials challenges for advanced reactor systems. *J. Nucl. Mater.* **385**, 217-222 (2009).

13. Samaras, M., Derlet, P.M., Van Swygenhoven, H. & Victoria, M. Atomic scale modelling of the primary damage state of irradiated fcc and bcc nanocrystalline metals. *J. Nucl. Mater.* **351**, 47-55 (2006).

14. Bai, X.-M., Voter, A.F., Hoagland, R.G., Nastasi, M. & Uberuaga, B.P. Efficient annealing of radiation damage near grain boundaries via interstitial emission. *Science* **327**, 1631-1634 (2010).

15. Bai, X.-M. *et al.* Role of atomic structure on grain boundary-defect interactions in Cu. *Phys. Rev. B* **85**, 214103 (2012).

16. Li, X.Y. *et al.* An energetic and kinetic perspective of the grain-boundary role in healing radiation damage in tungsten. *Nucl. Fusion.* **53**, 123014 (2013).

17. Sugio, K., Shimomura, Y. & de la Rubia, T.D. Computer simulation of displacement damage cascade formation near sigma 5 twist boundary in silver. *J. Phys. Soc. Jpn.* **67**, 882-889 (1998).

18. Pérez-Pérez, F.J. & Smith, R. Modelling radiation effects at GBs in bcc Fe. *Nucl. Instrum. Methods Phys. Res. B* **153**, 136-141 (1999).

19. Pérez-Pérez, F.J. & Smith, R. Structural changes at GBs in bcc Fe induced by atomic collisions. *Nucl. Instrum. Methods Phys. Res. B* **164-165**, 487-494 (2000).

20. Pérez-Pérez, F.J. & Smith, R. Preferential damage at symmetrical tilt GBs in bcc Fe. *Nucl. Instrum. Methods Phys. Res. B* **180**, 322-328 (2001).

21. Samaras, M., Derlet, P.M., Van Swygenhoven, H. & Victoria, M. Computer simulation of displacement cascades in nanocrystalline Ni. *Phys. Rev. Lett.* **88**, 125505 (2002).

22. Samaras, M., Derlet, P.M., Van Swygenhoven, H. & Victoria, M. Stacking fault tetrahedral formation in the neighbourhood of GBs. *Nucl. Instrum. Methods Phys. Res. B* **202**, 51-55 (2003).

23. Park, N.-Y. *et al.* Radiation damage in nano-crystalline tungsten: a molecular dynamics simulation. *Met. Mater. Int.* **15**, 447-452 (2009).

24. Li, X.Y. *et al.* Principal physical parameters characterizing the interactions between irradiation-induced point defects and several tilt symmetric GBs in Fe, Mo and W. *J. Nucl. Mater.* **444**, 229-236 (2014).

25. Yu, K.Y. *et al.* Removal of stacking-fault tetrahedral by twin boundaries in nanotwinned metals. *Nat. Commun.* **4**, 1377 (2013).

26. Frolov, T. *et al.* Structural phase transformations in metallic grain boundaries. *Nat. Commun.* **4**, 1899 (2013).

27. Fu, C.C., Torre, J.D., Willaime, F., Bocquet, J.-L. & Barbu, A. Multiscale modeling of defect kinetics in irradiated iron. *Nat. Mater.* **4**, 68-74 (2005).

28. Stoller, R.E., Kamenski , P.J. & Osetsky , Yu.N. Length-scale effects in cascade damage production in Fe. *MRS Proceedings* **1125**, 1125-R05-05 (2008).

29. Chen, D.,Wang, J., Chen, T. & Shao, L. Defect annihilation at grain boundaries in alpha-Fe. *Sci. Rep.* **3**, 1450 (2013).

30. Eshelby, J.D. The elastic interaction of point defects interaction. *Acta Metall.* **3**, 487 (1955).

31. Borovikov, V. *et al.* Coupled motion of GBs in bcc tungsten as a possible radiation-damage healing mechanism under fusion reactor conditions. *Nucl. Fusion.* **53**, 063001 (2013).

32. Galloway, G.J. & Ackland, G.J. Molecular dynamics and object kinetic Monte carlo study of radiation-induced motion of voids and He bubbles in bcc iron. *Phys. Rev. B* **87**, 104106 (2013).

33. Mendelev, M.I. *et al.* Development of new interatomic potentials appropriate for crystalline and liquid Fe. *Philos. Mag.* **83**, 3977-3994 (2003).

[1]  Derlet, P.M., Nguyen-Manh, D. & Dudarev, S.L. Multiscale modeling of crowdion and vacancy defects in body-centered-cubic transition metals. *Phys. Rev. B* **76**, 054107 (2007).

34. Ziegler, J.F., Ziegler, M.D. & Biersack, J.P. SRIM C The stopping and range of ions in matter. *Nucl. Instrum. Methods Phys. Res. B* **268**, 1818-1823 (2010).

35. Henkelman G. & Jónsson H. Improved tangent estimate in the nudged elastic band method for finding minimum energy paths and saddle points. *J. Chem. Phys.* 113, 9978–85 (2000).

**Acknowledgements**

This work was supported by the National Magnetic Confinement Fusion Program (Grant No.: 2011GB108004), the Strategic Priority Research Program of Chinese Academy of Sciences (Grant No.: XDA03010303), the National Natural Science Foundation of China (Nos.:91026002 and 11375231), and by the Center for Computation Science, Hefei Institutes of Physical Sciences.

**Author contributions**

C. S. Liu directed the entire study and discussed the results. Most of the calculations were performed by Xiangyan Li, Wei Liu and Yichun Xu. All authors contributed to the analysis and discussion of the data and the writing of the manuscript.

**Author Information** The authors declare no competing financial interests. Correspondence and requests for materials should be addressed to C. S. Liu. (csliu@issp.ac.cn).

Figure captions:

Figure 1

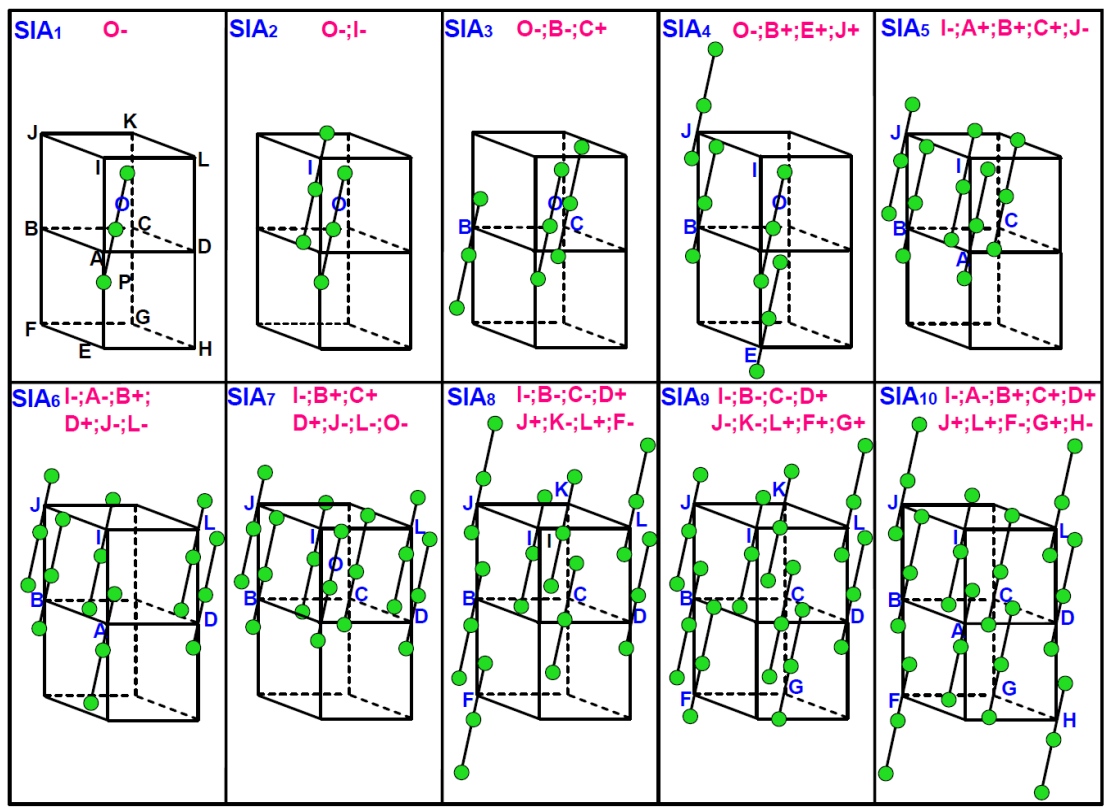

**Figure 1 | Stable configurations for small interstitial clusters in W bulk.** Here, $SIA_n$ indicates one interstitial cluster with $n$ single interstitials; $n$ ranges from 1 to 10. $A$, $B$, $C$, $D$, $E$, $F$, $G$, $H$, $I$, $J$, $K$, $L$, $O$, and $P$ mark the lattice sites in the two lattice cells. The letters marking the interstitial clusters are colored with blue. In the display scheme, three green spheres are aligned in the <111> direction, representing one crowdion. The symbol $O-$ marks the configuration of one $SIA$, meaning that it is created by initially inserting one atom at the position of $O-1/4$<111> and then relaxing the configuration. For $O+$, the atom is inserted at the position of $O+1/4$<111>. Other symbols marking the $SIA_n$ have similar meaning. The stable configuration of the $SIA_n$ is consisted of a bunch of <111> crowdions parallel to each other.

Figure 2

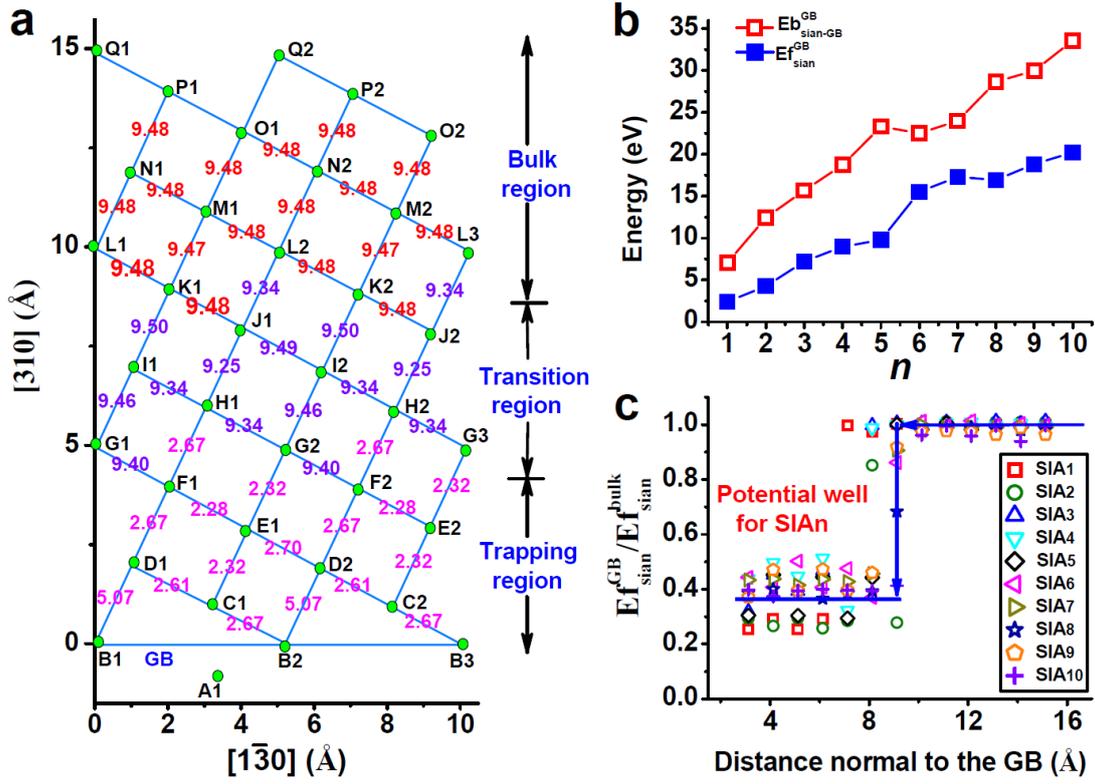

**Figure 2 | Energetics of $SIA_n$ near W symmetric tilt GB Σ5(3 1 0)/[0 0 1]. a**, Energy landscape for the single interstitial near the GB. Considering symmetry of the GB structure, results are shown just on one GB side. Interstitial formation energy is marked at the center of two nearest lattices that is the initial position of the interstitial before its relaxing. Line *B1-B3* indicates the GB position. According to the variation in the energy landscape, the spatial region near the GB is approximately divided into three parts: *Bulk region*, *Transition region*, and *Trapping region*. **b**, Formation energy for the $SIA_n$ at the GB ( $Ef_{sian}^{GB}$ ) and its binding energy with the GB ( $Eb_{sian-GB}^{GB}$ ). Here *n* is the number of interstitials in one interstitial cluster. **c**, Formation energy profile for $SIA_n$ near the GB. Here the formation energy is normalized by dividing respective bulk values ( $Ef_{sian}^{GB} / Ef_{sian}^{bulk}$ ).

Figure 3

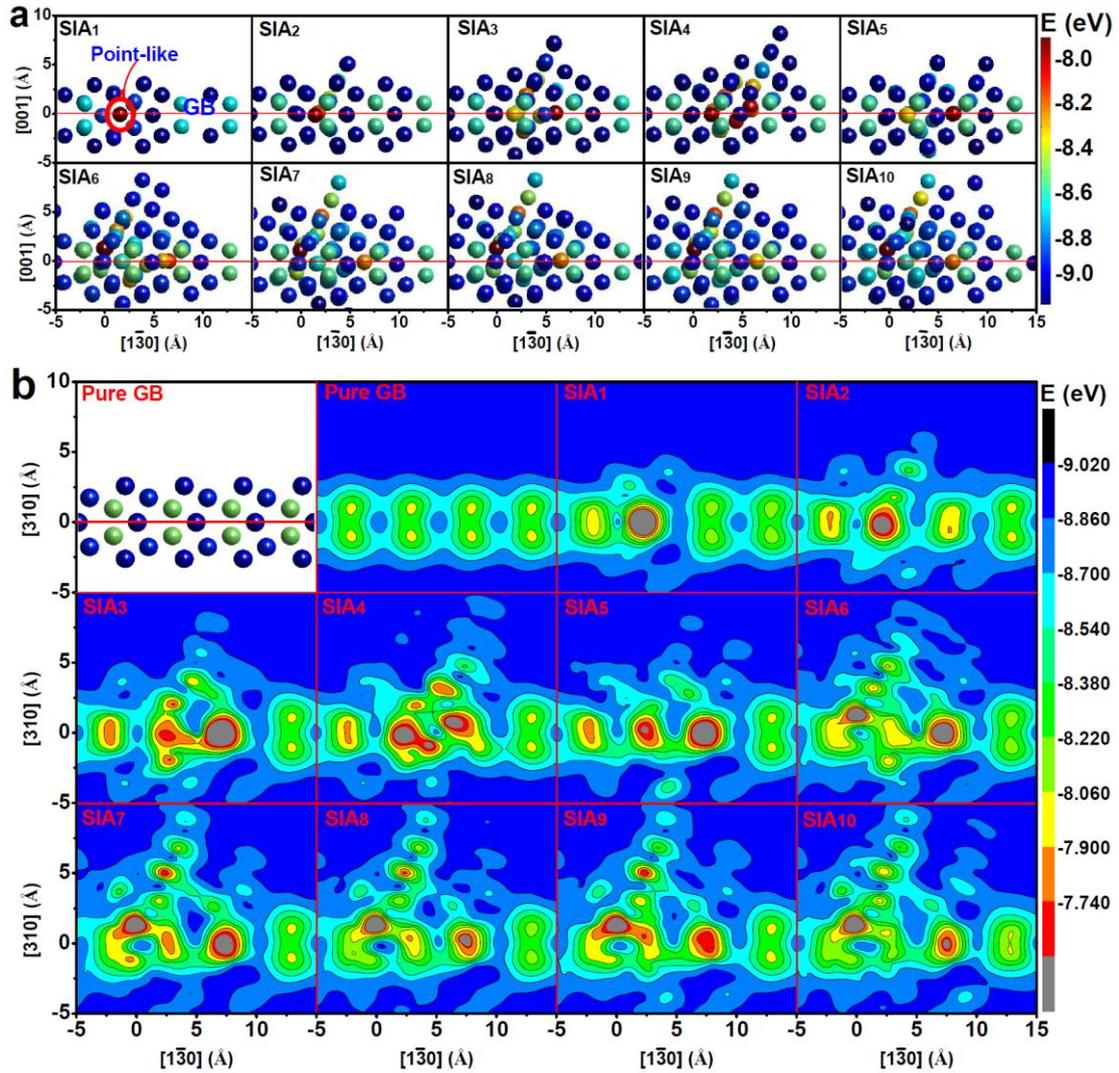

**Figure 3 | States for small interstitial clusters after their trapping at the GB in W. a**, Visualization of the *trapped state* for $SIA_1$–$SIA_{10}$ by atomic configurations near the GB. In this display scheme, atoms are colored with their potential energies; the atoms are not shown that have their potential energies deviation from bulk value less than 0.1 eV. The GB position is indicated by the red line. As one interstitial ($SIA_1$) or its clusters ($SIA_2$–$SIA_{10}$) are put near the pure GB, they instantly segregate into the GB and get trapped there during relaxing the GB structure. These interstitial clusters behave like point defects occupying the vacant space at the GB. **b,** Contour for atom potential energies near the GB that has trapped one interstitial cluster ($SIA_1$–$SIA_{10}$), corresponding to that in (**a**) as shown in real

space. For comparison, the atom configurations and their potential energy contour near the pristine GB are also presented.

Figure 4

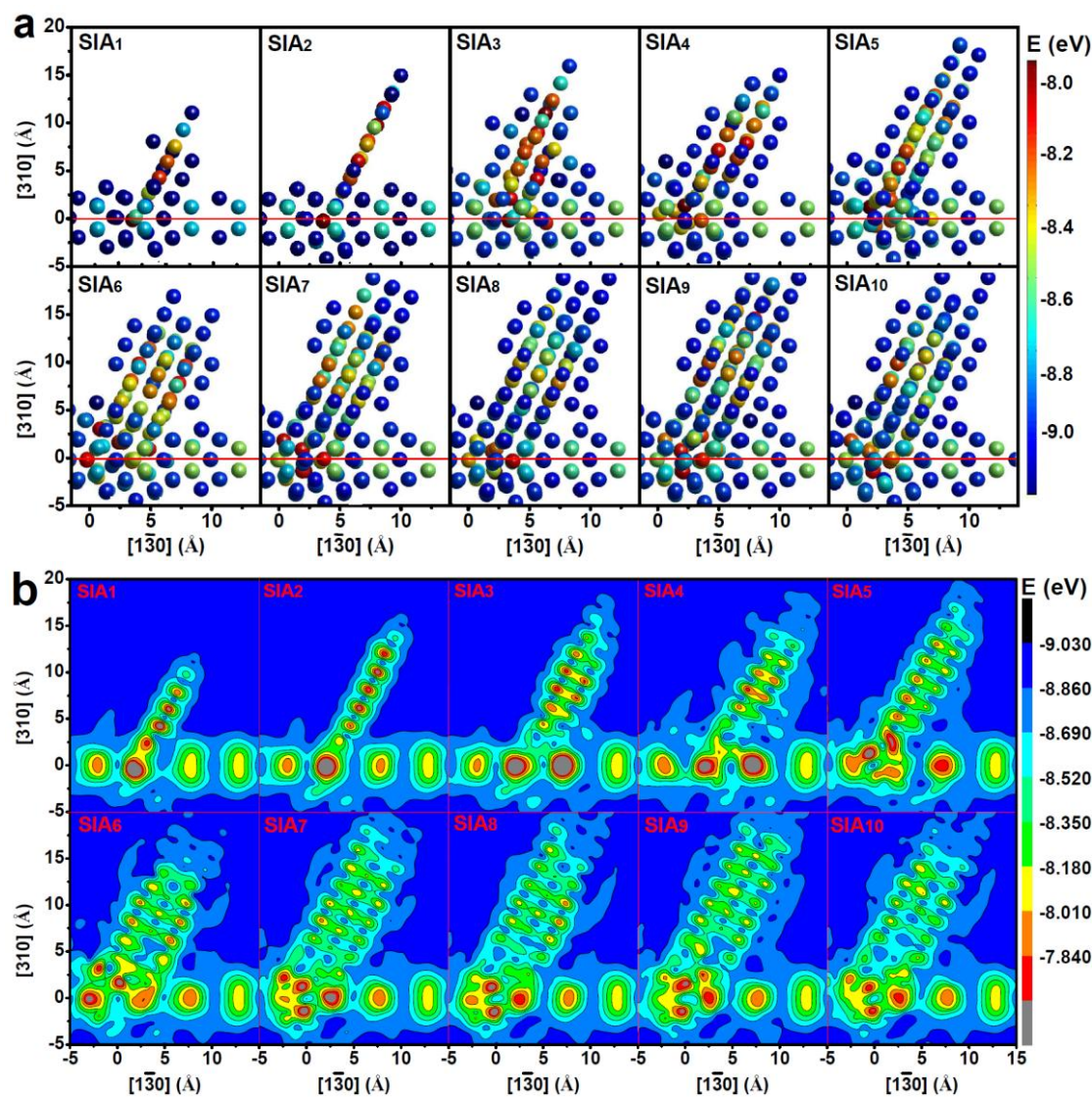

**Figure 4 | Blocking states for small interstitial clusters near the GB in W.**
**a**, Visualization of the *blocking state* for *SIA₁–SIA₁₀* in real space. The display scheme is identical to that in Fig. 3a. As more interstitials are created near the *loaded GB* (the GB that has trapped a number of interstitials like the one in Fig. 3), they no longer spontaneously flow into the GB, but get stuck in the vicinity of the *loaded GB*. These interstitial clusters basically maintain their shape of parallel crowdions as in the bulk.
**b**, Contour for atom potential energies near the GB with one interstitial cluster (*SIA₁–SIA₁₀*) stuck nearby, corresponding to that in **a**.

Figure 5

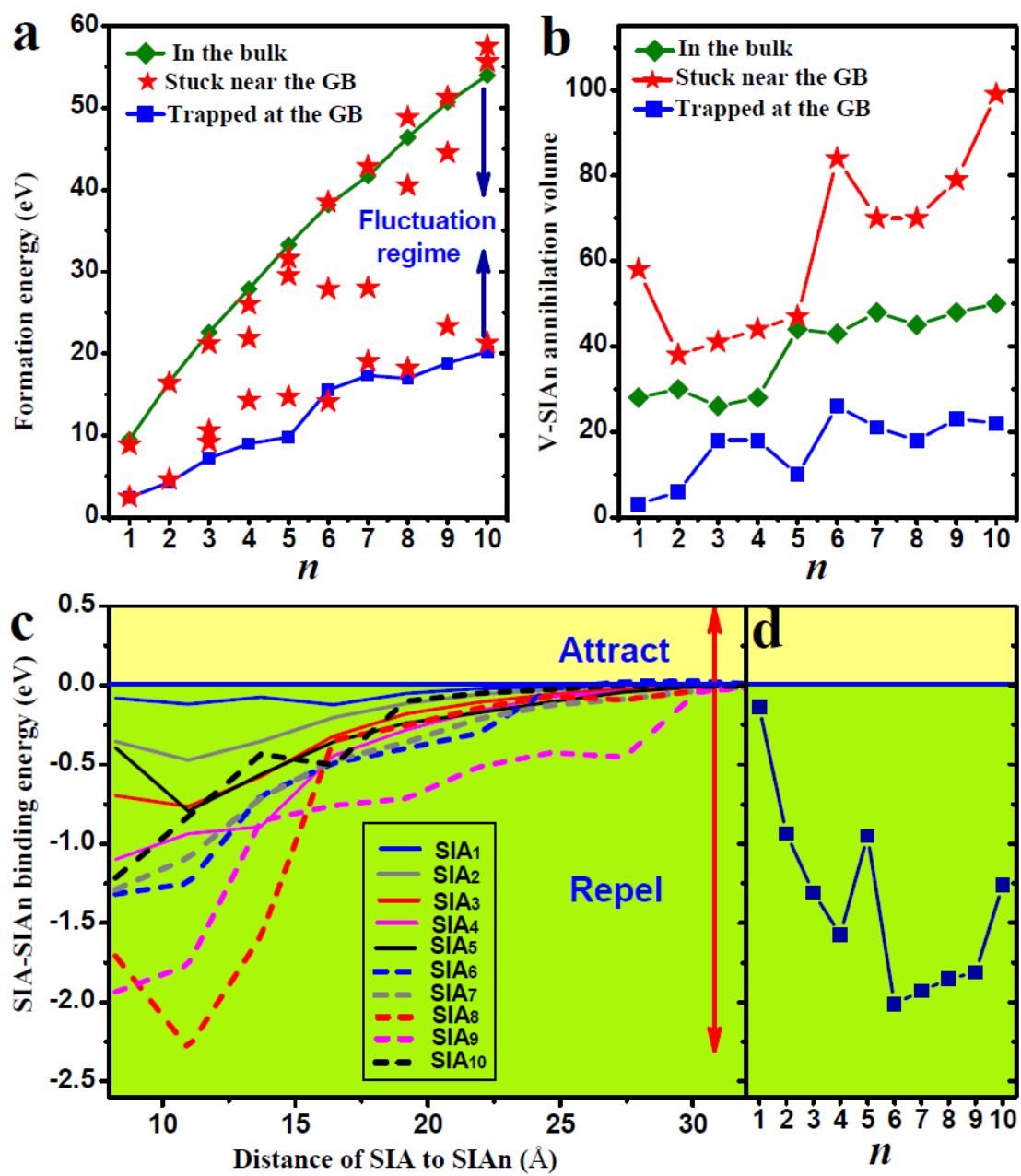

**Figure 5 | Energetic consequence of the blocking effect. a**, Formation energies for $SIA_1$–$SIA_{10}$ and **b**, annihilation volume for the vacancy around $SIA_1$–$SIA_{10}$ in the bulk, stuck near the GB, and trapped at the GB. The annihilation volume is calculated as the number of lattices around one interstitial cluster where a vacancy spontaneously recombines with one interstitial within the cluster. **c**, Binding energies of one interstitial to cluster that gets stuck near the GB and average binding energies of one interstitial to each interstitial within the cluster (**d**). The binding energies of

the interstitial to the cluster in are calculated in the direction along the long axis, since the interstitials stuck near the GB are orientated towards the bulk region.

Figure 6

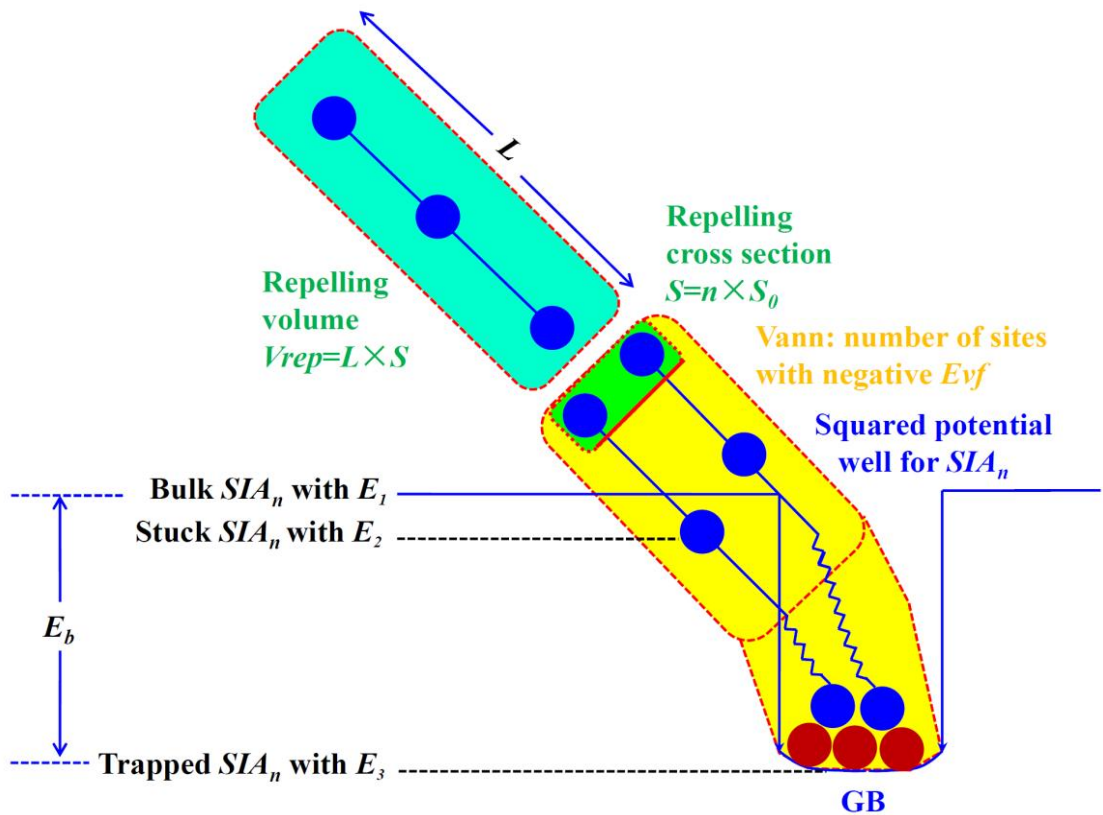

**Figure 6 | Illustration of interstitials states and their energy levels near the GB.** Three spheres lined together indicate one *SIA*. The GB acts as a squared potential well for the $SIA_n$ and trap interstitials as indicated by three red spheres at the GB. These interstitials located at the GB can further block interstitials near the GB, making them stuck there. The $SIA_n$ has rather different *energy levels* as it is in the bulk, stuck near the GB, and trapped at the GB. $E_1$ is $SIA_n$ formation energy in the bulk. $E_2$ and $E_3$ are formation energies of interstitials stuck near the GB and trapped at the GB, respectively. $E_b$ is binding energy of bulk $SIA_n$ to the GB, as given in Fig. 2b. As the $SIA_n$ is stuck near the GB, it has a significantly large annihilation volume for vacancies (*Vann*), as shown by the yellow region. *Evf* is vacancy formation energy around the $SIA_n$. Besides, the $SIA_n$ stuck near the GB owes one repelling cross section in the direction along <111>, as indicated by the green box. Repelling interaction extends along <111> direction as far as *L*, as indicated by the light blue box. The area

of the cross section for each *SIA* in the *SIA*$_n$ is $S_o = \frac{\sqrt{2}}{2} a^2$ with *a* as the lattice constant.

The area of the *SIA*$_n$ cross section is $S = n \times S_o$ with *n* as the number of *SIA* in one *SIA*$_n$. *L* is estimated to be 20 Å based on calculations in Fig. 5c. The repelling volume is *Vrep=L×S*.